\documentclass[12pt]{article}
\usepackage{graphicx}
\usepackage{amssymb,amsmath,amsfonts,amsthm}
\usepackage{amssymb}
\usepackage{epstopdf}
\usepackage[normalem]{ulem}
\usepackage{comment}
\newcommand{\beq}{\begin{equation}}
\newcommand{\eeq}{\end{equation}}

\newcommand{\f}{\begin{equation}}
\newcommand{\ff}{\end{equation}}

\usepackage{hyperref}

\hypersetup{
   colorlinks=true,         
   linkcolor=blue,          
   citecolor=red,        
   urlcolor=Violet            
}

\usepackage[usenames,dvipsnames]{color}

\let\oldmarginpar\marginpar
\renewcommand\marginpar[1]{\oldmarginpar{\color{red}\raggedright\scriptsize #1}}

\begin{document}

\title{Quantum energetic causal sets}
\author{Marina Cort\^{e}s${}^{1,2,3}$ and Lee Smolin${}^{1}$
\\
\\
Perimeter Institute for Theoretical Physics${}^{1}$\\
31 Caroline Street North, Waterloo, Ontario N2J 2Y5, Canada
\\
\\
Institute for Astronomy, University of Edinburgh ${}^{2}$\\
Blackford Hill, Edinburgh EH9 3HJ, United Kingdom
\\
\\
Centro de Astronomia e Astrof\'{i}sica da Universidade de Lisboa${}^{3}$\\
Faculdade de Ci\^encias, Edif\'{i}cio C8, Campo Grande, 1769-016 Lisboa, Portugal}

\date{\today}

\maketitle

\begin{abstract}
We propose an approach to quantum theory based on the energetic causal sets, introduced in~\cite{first}. 
Fundamental processes are causal sets whose events carry  momentum and energy, which are transmitted along causal links and conserved
at each event. Energetic causal set models differ from other spacetime-free causal set approaches, e.g.~Ref.~\cite{fotini} proposed causal sets based on quantum information processing systems, and Ref.~\cite{cohl} proposed causal sets constructed out of standard model particles.
Fundamentally there are amplitudes for causal processes in energetic causal sets, but no space-time.  
An embedding of the causal processes
in an emergent space-time arises only at the semiclassical level.  Hence, fundamentally there are no commutation relations, no uncertainty principle and, indeed, no $\hbar$.  All that remains of quantum theory is the relationship between the absolute value squared of complex amplitudes and probabilities.  Consequently, we find that neither locality, nor non locality, are primary concepts, only causality exists at the fundamental level.
\end{abstract}

\newpage

\tableofcontents

\section{Introduction}


Relativistic quantum mechanics is sometimes taken to be just a sub case of quantum mechanics, with the same postulates - and consequently the same resulting
foundational issues - as  its non-relativistic versions. 
In particular, the standard formulations of quantum field theory have all the usual furniture of quantum theory including non-commutative operator algebras and their Hilbert space representatives.

Here  we propose here a novel foundation for relativistic quantum theory in which operators play no role.  We posit that at the fundamental level
there is only causality, energy and momentum,  while space-time is not part of the fundamental description of nature.
Consequently, there are no operator algebras because there is nothing for the momentum and energy to fail to commute with.  The part of quantum theory that we {\it do} need to postulate is the superposition principle, as well as the interpretation of the probability as square of complex amplitudes of individual processes.  Space-time emerges only in the classical limit, as does the quantum theories  of  free and interacting relativistic particles moving in that space-time.
 
In this paper, 
we are concerned with calculating
the probabilities for total processes, in which a set of incoming particles, specified by their energy-momenta (below we will use momenta to refer
to energy-momenta unless otherwise specified) and perhaps other quantum numbers, are converted to a set of outgoing particles with specified
momenta.   As in conventional quantum theory we assume (see the precise formulation below) that these probabilities are the square of amplitudes. These amplitudes are complex
numbers, and we  assume that the amplitude for the total processes is gotten by summing up amplitudes for elementary processes by
means of which the transition may occur.   
Apart from that we assume no other ingredients of quantum theory: no space-time variables, hence no uncertainty relations or commutation relations 
and no Hilbert space of states.  Indeed $\hbar$ is not mentioned in the formulation of the theory and arises only when and if space-time coordinates emerge at a classical level.  Rather than being fundamental, $\hbar$ is an artifact of convention and appears only if we insist on measuring positions
in units of length rather than  in units of inverse momenta.   

The processes that we posit as fundamental have two kinds of structures.  The first is causal structure: we assume an ontology of discrete events related by causal relations.  Each process is a causal set which expresses a relational view of fundamental physics in which the identity of an event is defined by its causal relations to other events.  This approach is in line with the causal set program~\cite{cs}, and see~\cite{cs2} for a review. However we propose an addition to the causal set: we do not stick to a purely relational description and ascribe to each event intrinsic energy and momenta. These are transmitted by the causal links and have conservation laws at each event.  A causal set so decorated may be called an {\it energetic causal set.} The theory we discuss here can be called {\it quantum energetic causal sets.}

Adding an intrinsic quantity such as momenta to an otherwise purely relational causal set may be motivated in three ways:  

\begin{enumerate}

\item{} It can be argued that the world cannot be purely defined by relationships.  Relationships must relate something. If elements in a relational set don't have labels, or an associated quantity, there will be no way of specifying which events are related. The events that make up the world must have an intrinsic quantity that allows them to be related to each other.

\item{}The formulation of relative locality~\cite{rl,rl2} teaches us that momentum and energy are the fundamental observables of dynamics.  Space-time is a conventional construction, defined operationally, as Einstein taught us, by sending and receiving quanta that carry energy and momenta.

\item{}A major issue with the causal set program is getting space-time to emerge from a causal set.  This problem is solved by the construction of energetic causal sets, as we show below.

\end{enumerate}

Another issue that is addressed in this formulation is non-locality. 
Since space-time is emergent, at the fundamental level there is neither locality nor non-locality, just causality.

Before closing this section we may note that some examples of quantum energetic causal sets are provided by
{\it quantum causal histories}\cite{QCH}.  Here there is a Hilbert space associated with the causal propagation of free particles, ${\cal H}$. In the case of a relativistic theory that is the Hilbert space of free relativistic particles, spanned by momentum eigenstates.   An interaction may be described by an evolution operator, which is a completely positive map,  defined on the product of Hilbert spaces of the incoming particle, so physical states are functionals $\psi ( p )  $ of the mass shell.  In this case there automatically are position operators defined by 
$\hat{x}^a = \frac{\hbar}{\imath} \frac{\partial}{\partial p_a}$. 

However two points need to be emphasized.

First, in quantum causal histories one does not sum over  causal structures as one does here. So the two constructions are different and the position operators just mentioned are not directly related  to the embedding coordinates $z^a$ of events to be define below.

Second, this does not imply that every quantum energetic causal set, as defined below, comes from such a construction with a fundamental Hilbert space.  Nor does it mean that the Hilbert spaces need be taken as fundamental, even in the cases where they may occur.  We leave this question open to be resolved by future work.  

This paper is the second in a series investigating subjects related to causality and irreversibility in fundamental physics.  Energetic causal sets were
introduced in the first paper~\cite{first}; this paper revisits some of the results presented there in the context of quantum dynamics, rather than the deterministic dynamics introduced earlier.  
In a related paper\cite{ECS-3}, we show that a class of spin foam models recently proposed by Wieland\cite{WW} can be understood to be energetic causal sets.
Irreversibility does not play a large role in the quantum theory we derive here since we interpret it as an effective theory (although see Section~\ref{conclusions}) but is the focus of the first, and will be that of a later paper in this series~\cite{motivation}. The work in this series is also related to the program set out in~\cite{TR,SU}.  

\section{Basic principles}

We wish to study an isolated process $\cal S$ which is specified by a list of input particles with energy-momentum $p_a^{in, \ I}$
and outgoing particles with energy-momenta $q_a^{out,\  I}$.  These energy-momenta live on a $d-$dimensional momentum space
$\cal P$, which is endowed with a metric $h^{ab}$ and connection $\Gamma_a^{bc}$.  In the spirit of relative locality the geometry
of $\cal M$ can be assumed to be general, but in this paper when we need to be specific we will assume it is Minkowskian.

The whole quantum process specified by just the incoming and outgoing particles and momenta will be called the total process.
We will posit the usual probability rule for quantum mechanics specified by the following postulates:

\begin{itemize}

\item{}Corresponding to a given total process there are an infinite number of elementary processes.

\item{}An elementary process is a labeled causal set whose events are elementary events.  Each elementary event converts a set of incoming particles  into a set of outgoing particles; all particles are labeled by momentum that live in $\cal P$.   
 $p_{aK}^I$ is the momenta incoming to event $I$ from event $K$ and $q_{aI}^L$ is the momenta outgoing from event $I$ to event $L$.

\item{} The dynamics is specified by proscribing to each elementary event, $I$, a complex amplitude, ${\cal A}_I$. 

\item{}The {\it amplitude} for an elementary process, ${\cal A}[P]$, is the product of amplitudes at each event.  
\f
{\cal A}[P] = \prod_I {\cal A}_I
\ff
 
 \item{}The amplitude for a total process labeled by incoming and outgoing particles and their momenta is the sum of the amplitudes
 for the elementary processes that have those incoming and outgoing particles.
 \f
 {\cal A}[p_a^{in, \ I};  q_a^{out,\  I}  ] = \sum_P {\cal A}[P]
 \ff
 
 \item{}The probability for the total processes is the absolute value squared of the total amplitude
\f
 {\cal P}[p_a^{in, \ I};  q_a^{out,\  I}  ] = \left |  {\cal A}[p_a^{in, \ I};  q_a^{out,\  I}  ] \right |^2  
\ff

\item{}It follows from the definition of probability that the amplitudes must be chosen subject to the constraint that the probabilities are normalized so that
\f
\sum_{q_a^{out,\  I} } {\cal P}[p_a^{in, \ I};  q_a^{out,\  I}  ] = \sum_{q_a^{in,\  I} } {\cal P}[p_a^{in, \ I};  q_a^{out,\  I}  ] =1
\label{normalized}
\ff

\end{itemize}

These postulates together with the rule that assigns amplitudes to events completely specifies the theory.  

At this level there
are no space-time variables, hence no commutation relations, hence no uncertainty relations.  There is indeed no $\hbar$, so it
would be impossible to write commutation relations or uncertainty relations.  There is no Hilbert space of states.

The momenta are subject to three sets of constraints.
\begin{enumerate}

\item{}Conservation laws:
\f
{\cal P}_a^I = \sum_K p_{a K}^I -  \sum_L q_{a I}^L  =0 
\ff
where the sum over $K$ is over all events $I$ is connected to in the past and the sum over $L$ is over all events $I$  is connected to in the future.
\item{} No redshifts
\f
{\cal R}_{aI}^K = p_{aI}^K - q_{aI}^K =0 
\ff
\item{}Energy momentum relations for relativistic particles
\f
{\cal C}^I_K = \frac{1}{2} \eta^{ab} p_{a K}^I p_{b K}^I  + m^2 =  0 , \ \ \ \ \  \tilde{\cal C}^I_K = \frac{1}{2} \eta^{ab} q_{a K}^I q_{b K}^I  + m^2=  0 
\label{onshell}
\ff

\end{enumerate}

\section{The emergence of space-time}\label{qm_st}

\subsection{Introduction of path integrals}

To see how space-time emerges from these principles we write the amplitude for an elementary process ${\cal A}[P]$ as a path integral,
\f
{\cal A}[P]  =   \int \prod_{IJ} dp_a^{IJ} dq_a^{IJ} \delta ({\cal C}_a^{IJ} ) \delta ({\cal R}_I^J ) \prod_{I }  \delta ({\cal P}_a^{I} ) 
\prod_I {\cal A}_I 
\label{quantum0}
\ff
where the delta functions impose the constraints.  
Note that we integrate over all the internal momenta leaving the incoming and outgoing momenta fixed.

We next introduce three sets of lagrange multipliers to exponentiate the three constraints constraints.  
\f
{\cal A}[P] =  N [ {\cal C}] \int \prod_{IJ} dp_a^{IJ}  dq_a^{IJ} d{\cal N}_I^J  d \tilde{\cal N}_I^J  \prod_{I } dZ^a_I  \ e^{\imath S^0} 
\label{quantum1}
\ff
where the dimensionless action is
 \f
 S^0= \sum_I Z^a_I {\cal P}_a^I   +\sum_{(I,K)} (   X^{a I}_K {\cal R}_{aI}^K + {\cal N}^K_I {\cal C}^I_K +  \tilde{\cal N}^K_I  \tilde {\cal C}^I_K )
 + S^{int}
 \label{S0}
 \ff
where the sum over $(I,K)$ is over all connected pairs of events.  
There are three kinds of lagrange multipliers, $Z^a_I$ associated with each event, $X^{a K}_I$ and 
 the ${\cal N}^I_K$ and  $\tilde{\cal N}^I_K$ are associated with each connected pair of events. 
 Note that the $Z^a_I$ and the $X^{a K}_I$ have dimensions of inverse momenta.
 
 The interaction action, $S^{int}$ is given by
 \f
 S^{int} = -\imath \ln \prod_I {\cal A}_I   
 \label{Sint}
 \ff

Note that  the normalization
constant $N [ {\cal C}] $ depends on the causal structure.

\subsection{The emergence of space-time}

We can now show that the $Z^a_I$ play the role of emergent space-time coordinates of the events.  To see this we consider the stationary phase
approximation to the path integral.  We will first discuss the case that the particles are all massless, so that the constraints (\ref{onshell}) imply the momenta are
null vectors in momentum space, $\cal P$.  

Let us  assume  that the products $Z \cdot p$ in the action (\ref{S0}) are large compared to unity so that we can evaluate the integrals in
(\ref{quantum1}) in the stationary phase approximation.  We then seek the critical points of $S^0$.  We will see that this leads to the emergence of space-time.  

The variation of the action by the lagrange multipliers gives the constraints.  But we have new equations satisfied by the lagrange multipliers coming from the variation of the action by the momenta.  
\f
\frac{\delta S^0}{\delta p_{a K}^I }= Z^a_I + X^{a K}_I + {\cal N} p^{a I}_K =0 
\ff
\f
\frac{\delta S^0}{\delta q_{a  I }^K } = -Z^a_K - X^{aK}_I +  \tilde{\cal N} q^{a K}_I =0 
\ff
Adding these two equations and using $ {\cal R}_I^K =0 $ we find
\f
 Z^a_I -  Z^a_K =  p^{a I}_K (\tilde{\cal N}_I^K  + {\cal N}_I^K )
 \label{emergence}
\ff
This has a simple physical interpretation.  The lagrange multiplier $Z^a_I$ can now be interpreted as the  space-time coordinate of the
event $I$-expressed in units of inverse homenta.  $ Z^a_I -  Z^a_K $ is then a space-time interval between event $K$ and event $I$.  It is a light-like interval proportional to
the momentum $p^{a I}_K$ connecting $K$ to $I$.  The constant of proportionality involves the lagrange multipliers  $\tilde{\cal N} + {\cal N} $ which is consistent with the fact that the affine parameter along a null ray is arbitrary.  

We may note that equations (\ref{emergence}) will not always have simultaneous solutions.  There are one equation to be solved for every causal link, but only one $Z_I^a$ for each event.  In the cases where the equations can always be solved it means that there are a consistent choice of embeddings
$Z_I^a$ of the events in a flat space-time such that the causal links are represented by null intervals proportional to the energy-momentum they carry.
In these cases, 
we can say that {\it space-time has emerged,}  as there was no space-time and no locating the events in space-time in the original description.  It is interesting to note that the emergent space-time inherits the metric $\eta^{ab}$ from momentum space. 

The lagrange multipliers $Z^a_I$ start off as just arbitrary variables with no physical meaning other than reinforcing the constraint.  In the stationary phase approximation in which the action is an extremum under all variations, the $Z^a_I$ become coordinates embedding the events in
Minkowski spacetime-where the metric of space-time comes from momentum space.  Apart from the classical limit which is defined by the stationary phase approximation,  the notion of  space-time has no meaning. 

It is desirable to measure the space-time coordinates in units of length. To do this we must introduce a constant $\hbar$ with units  of action
so that
\f
z^a_I = \hbar Z^a_I
\ff
has dimensions of length.  Note that $\hbar$ has no fundamental meaning, it is purely conventional and arises from a nostalgic desire to measure space-time coordinates in units of length.  

In the case that the particles are massive the intervals $ Z^a_I -  Z^a_K$ between the event's coordinates defined by eq. (\ref{emergence})
are timeline rather than null, but the picture is similar.

\subsection{The emergence of relativistic particle dynamics from chain of events}

We can consider the example of a chain of events ${\cal E}_I$, $I=1,É N$, which each have a single incoming and single outgoing momenta, denoted simply by $p_a^I $ and $q_{a I}$.  
Alternatively, we can regard these as a chain of interactions at which all but one incoming and one outgoing momenta are negligible.

We can solve the $\cal R$ constraints to find 
\f
p_a^I = q_a^I 
\ff
We then have for the exponentiated ${\cal P}_a^I$ constraints,
\f
S^0= \sum_I Z^a_I {\cal P}_a^I + {\cal M}_I {\cal C}^I
\ff
where 
\f
{\cal M}_I =  \tilde{\cal N}_I^{I-1} + {\cal N}^{I}_{I+1}
\ff
The first term can be expanded,
\f
\sum_I Z^a_I {\cal P}_a^I = \sum_I Z^a_I ( p^I_a - p^{I-1}_I ) \approx \sum_I Z^a_I \dot{p}_a^I \Delta t
\ff
where $\Delta t$ is a small interval and $ p^I_{\alpha} -p^{I-1}_a \approx \dot{p}_a (t) \Delta t $

If we take the limit of $\Delta t \rightarrow 0$ so that $\sum_I \Delta t \rightarrow \int dt$ so that the chain goes over in the limit to a curve.
In this limit the $p_a^I$ can be replaced by the continuous functions $p_a(t)$.  We will also replace the
discrete $Z_I^a$ by continuous variables $z^a (t)$ so that
the action becomes
\f
S^{free} = \int dt \left ( p_a (t) \dot{z}^a (t)  - \frac{1}{2}  n (t)  ( p_I^2 +m^2) \right )
\label{Sfree}
\ff
where $n= \frac{{\cal M}_I}{\Delta t} $ remains finite as both $\Delta t $ and  ${\cal M}_I$ are taken to zero.
(\ref{Sfree})  is the action for a free relativistic particle.  

Note that this form is invariant under reparameterizations 
\f
t \rightarrow t^\prime =  f(t) , \ \ \ \ \  dt^\prime = \dot{f} dt ,\ \ \ \ n (t) \rightarrow n^\prime (t^\prime ) = \frac{n(t)}{\dot{f}}
\ff
$n(t)$ is a lagrange multiplier that gives the energy-momentum relation as a constraint, $p^2+m^2 =0$

The $z^a (t) $ are then lagrange multipliers that enforce the equation of motion for a free particle
\f
\dot{p}_a (t)\ =0
\ff
while the variation by $p_a (t)$ gives the relation between the velocity and momenta
\f
\dot{z}^a = n (t) p^a
\ff
where the lagrange multiplier is arbitrary reflecting the reparameterization invariance.

Next we consider a network of long chains connected together by intersections of three or more chains.  We continue to label these 
intersections by $I$ and the chains that connect them by $(I,J)$.  Taking again the limit of $S^0$ for each chain we
have
\f
S^0 \rightarrow S^{rel} = \sum_{I,J} S^{free}_{I,J} + \sum_I z^a_I {\cal P}^I_a  ,
\label{Srel}
\ff
where the conservation law ${\cal P}^I_a$ is a function of the momenta at the endpoints of the paths that meet at the intersection
point $I$.

$S^{rel}$ given by (\ref{Srel})  is the action for a process in which a set of free relativistic particles interact at intersections where the conservation laws
${\cal P}^I_a$ are satisfied.  Note that the equations of motion for the $p_a(s)$ at the end points enforce the locality of the interactions by
equating $z^a_I$,  the coordinate of the $I$'th intersection with the coordinate of the endpoint of the path that meet there,
$z^a (s=1) $ or $z^a (s=0) $.

\section{Conclusions}\label{conclusions}

This is the second of a series of papers focused on studying the role of irreversibility in fundamental physics \cite{first,ECS-3,motivation}.
Here we presented a novel approach to quantum theory based on energetic causal sets introduced in the first work.  

The quantum theory we obtain, which we call {\it quantum energetic causal sets}, requires only a reduced set of the usual postulates of quantum mechanics. 
This is possible because our theory takes energy-momentum, instead of space-time, as fundamental. We only obtain space-time and its geometry as emergent properties of the semi-classical limit. 
This is in accord with the lesson
Einstein taught us: the concepts of simultaneity and locality are constructed from primary observations of energy and momentum. This is also the main result of relative locality: \textit{the primary geometry
is the geometry of energy-momentum space}.
This way,  energy-momentum and causality are the only fundamental intrinsic
properties of quantum histories. 
Taking energy-momentum as fundamental has the additional advantage that if space-time is emergent then so is locality. Fundamentally there is neither locality nor non-locality, \textit{just causality}.   

The quantum theory we obtain has, at the fundamental level, no non-commutativity, no uncertainty relations, and no $\hbar$. These all emerge with space-time.
The only postulate we require of the quantum theory is the amplitude law. 

We close with some comments on open questions.

\begin{itemize}

\item{}  As we mentioned in the introduction, there are examples of quantum energetic causal sets where the amplitudes can be defined by choices of Hilbert spaces and operators assigned to the causal links and events\cite{QCH}.  But this does not conflict with our assertion that these are a set of quantum theories in which space-time and associated notions, including the canonical commutation relations, the uncertainty principle and the value of $\hbar$ are emergent rather than fundamental.

\item{}It will be interesting to understand whether a Hilbert space formulation of quantum theory can always  be reconstructed for every quantum energetic causal set or, if not, when that can be done.

\item{}Spinors  can be incorporated naturally in the construction of an energetic causal set as shown in \cite{first,ECS-3}.
There is also no obstacle to adding other intrinsic labels such as discrete spin values to the intrinsic momenta.  Whether its possible to derive the spectrum of angular momentum or any other observable from the postulates used here is an important research question, but one beyond the scope of this paper.   

\item{} It is important to remember  that the stationary phase condition is a necessary but not sufficient condition for a history in a path integral to dominate the classical limit of a quantum process.  In ordinary quantum mechanics, and potentially in our case as well, there can be many stationary points, but only one may dominate the integral.  To understand this in more detail will require more work that is beyond the scope of this paper.

\item{} In future work we wish to investigate the hypothesis that the irreversible processes of the energetic causal set in~\cite{first} which are absent here, in the partial formulation of quantum mechanics, might play a role in the origin of the superposition of processes. This would explain the complex numbers in quantum mechanics amplitudes, and would allows us to reconstruct quantum theory from more natural postulates.

\end{itemize}
 
\section*{Acknowledgements}

 We are grateful to Jacob Barnett,  Astrid Eichorn, Fay Dowker,  Laurent Freidel, Raphael Sorkin,  Sumati Surya and a referee for conversations and encouragement. 
This research was supported in part by Perimeter Institute for Theoretical Physics. Research at Perimeter Institute is supported by the Government of Canada through Industry Canada and by the Province of Ontario through the Ministry of Research and Innovation. This research was also partly supported by grants from NSERC, FQXi and the John Templeton Foundation.
M.C.\ was supported by the EU FP7 grant PIIF-GA-2011-300606 and FCT grant SFRH/BPD/79284/201(Portugal).  


\end{document}